\documentclass[conference]{IEEEtran}

\IEEEoverridecommandlockouts

\usepackage[textsize=tiny,disable]{todonotes} 

\usepackage{pdflscape}
\usepackage{lscape}
\usepackage{rotating}
\usepackage{afterpage}

\newcommand{\mv}[1]{{\todo[inline,color=green!20]{Mayank: #1}}}
\newcommand{\ben}[1]{{\todo[inline,color=blue!20]{Ben: #1}}}

\newcommand{\emily}[1]{{\todo[inline,color=red!20]{Emily: #1}}}
\newcommand{\ay}[1]{{\todo[inline,color=yellow!20]{Arkady: #1}}}
\newcommand{\discuss}[1]{{\todo[inline,color=orange!20]{Discuss: #1}}}

\usepackage[cmex10]{amsmath}
\usepackage{amsthm}
\usepackage{array}
\usepackage{algorithm}
\usepackage{algpseudocode}
\usepackage{hyperref}
\theoremstyle{plain}

\theoremstyle{definition}

\usepackage{xspace}

\newcommand{\Init}{\ensuremath{\mathbf{Init}}\xspace}
\newcommand{\Query}{\ensuremath{\mathbf{Query}}\xspace}
\newcommand{\Insert}{\ensuremath{\mathbf{Insert}}\xspace}

\newcommand{\Update}{\ensuremath{\mathbf{Update}}\xspace}
\newcommand{\Rebuild}{\ensuremath{\mathbf{Refresh}}\xspace}

\newcommand{\server}{\ensuremath{S}\xspace}

\newcommand{\querier}{\ensuremath{Q}\xspace}

\newcommand{\oram}{Oblivious Index\xspace}
\newcommand{\shortoram}{\ensuremath{\mathtt{Obliv}}\xspace}
\newcommand{\sse}{Custom Index\xspace}
\newcommand{\shortsse}{\ensuremath{\mathtt{Custom}}\xspace}
\newcommand{\ppe}{Legacy Index\xspace}
\newcommand{\shortppe}{\ensuremath{\mathtt{Legacy}}\xspace}

\definecolor{stealthS}{rgb}{0,1,0}
\definecolor{stealthP}{rgb}{1,1,0}
\definecolor{bsS}{rgb}{.25,.75,0}
\definecolor{bsP}{rgb}{.4,.6,0}
\definecolor{espadaS}{rgb}{.35,.65,0}
\definecolor{espadaP}{rgb}{.25,.75,0}
\definecolor{arxS}{rgb}{.5,.5,0}
\definecolor{arxP}{rgb}{.25,.75,0}
\definecolor{cryptdbS}{rgb}{.75,.25,0}
\definecolor{cryptdbP}{rgb}{0,1,0}

\usepackage{mdwlist} 
\usepackage{cite}    
\usepackage{url}
\usepackage{multirow}
\usepackage{tikz}

\usepackage{adjustbox,booktabs,multirow} 
\usepackage{graphicx}    
\usepackage{colortbl}

\newcommand*\rot{\rotatebox{90}}                                   
\newcommand{\rots}[2]{\parbox[t]{2mm}{\multirow{#1}{*}{\rot{#2}}}} 

\usepackage[nointegrals]{wasysym}
\newcommand{\full}{\CIRCLE\xspace}
\newcommand{\half}{\LEFTcircle\xspace}
\newcommand{\none}{\Circle\xspace}
\newcommand{\oneQuarter}{\ensuremath{\circleurquadblack}\xspace}
\newcommand{\threeQuarter}{\ensuremath{\blackcircleulquadwhite}\xspace}

\usepackage{stix} 

\newcommand{\nonePie}{\none}
\newcommand{\halfPie}{\half}
\newcommand{\fullPie}{\full}

\usepackage{pifont}
\newcommand{\yes}{\ding{52}}

\newcommand{\maybeCircle}{\fakeHalf}

\newcommand{\fakeHalf}{\half}

\newcommand{\secref}[1]{\mbox{Section~\ref{#1}}}

\newcommand{\apref}[1]{\mbox{Appendix~\ref{#1}}}

\newcommand{\tabref}[1]{\mbox{Table~\ref{#1}}}

 \newcommand{\ifblinded}[2]{#2} 

\begin{document}


\title{SoK: Cryptographically Protected Database Search\thanks{This material is based upon work supported under Air Force Contract No.~FA8721-05-C-0002 and/or FA8702-15-D-0001. Any opinions, findings, conclusions or recommendations expressed in this material are those of the author(s) and do not necessarily reflect the views of the U.S.~Air Force. The work of B.~Fuller was performed in part while at MIT Lincoln Laboratory. The work of M.~Varia was performed under NSF Grant No.~1414119 and additionally while a consultant at MIT Lincoln Laboratory.}}

\author{\IEEEauthorblockN{
Benjamin Fuller\IEEEauthorrefmark{1}, Mayank Varia\IEEEauthorrefmark{2}, Arkady Yerukhimovich\IEEEauthorrefmark{3}, Emily Shen\IEEEauthorrefmark{3}, Ariel Hamlin\IEEEauthorrefmark{3}, \\Vijay Gadepally\IEEEauthorrefmark{3}, Richard Shay\IEEEauthorrefmark{3}, John Darby Mitchell\IEEEauthorrefmark{3}, and Robert K.~Cunningham\IEEEauthorrefmark{3}}
\vspace{.04in}
\IEEEauthorblockA{\IEEEauthorrefmark{1}University of Connecticut\\
Email: benjamin.fuller@uconn.edu}
\vspace{.04in}
\IEEEauthorblockA{\IEEEauthorrefmark{2}Boston University\\
Email: varia@bu.edu}
\vspace{.04in}
\IEEEauthorblockA{\IEEEauthorrefmark{3}MIT Lincoln Laboratory\\
Email: \{arkady, emily.shen, ariel.hamlin, vijayg, richard.shay, mitchelljd, rkc\}@ll.mit.edu}
}



%

\maketitle



\begin{abstract}
Protected database search systems cryptographically isolate the roles of reading from, writing to, and administering the database.  This separation limits unnecessary administrator access and protects data in the case of system breaches. Since protected search was introduced in 2000, the area has grown rapidly; systems are offered by academia, start-ups, and established companies.

However, there is no best protected search system or set of techniques.  Design of such systems is a balancing act between security, functionality, performance, and usability.  This challenge is made more difficult by ongoing database specialization, as some users will want the functionality of SQL, NoSQL, or NewSQL databases.  This database evolution will continue, and the protected search community should be able to quickly provide functionality consistent with newly invented databases.

At the same time, the community must accurately and clearly characterize the tradeoffs between different approaches.  To address these challenges, we provide the following contributions:
\begin{enumerate}
\item An identification of the important primitive operations across database paradigms.  We find there are a small number of \emph{base} operations that can be used and combined to support a large number of database paradigms.
\item An evaluation of the current state of protected search systems in implementing these base operations. This evaluation describes the main approaches and tradeoffs for each base operation.  Furthermore, it puts protected search in the context of unprotected search, identifying key gaps in functionality.
\item An analysis of attacks against protected search for different base queries.
\item A roadmap and tools for transforming a protected search system into a protected database, including an open-source performance evaluation platform and initial user opinions of protected search.
\end{enumerate}
\end{abstract}



\section{Introduction}
The importance of collecting, storing, and sharing data is widely recognized by governments~\cite{Powers2014}, companies~\cite{Linoff:2002:MWT:560274,insightdata}, and individuals~\cite{Mons2011}.  When these are done properly, tremendous value can be extracted from data, enabling better decisions, improved health, economic growth, and the creation of entire industries and capabilities.

Important and sensitive data 
are stored in database management systems (DBMSs), which support ingest, storage, search, and retrieval, among other functionality.
DBMSs are vital to most businesses and are used for many different purposes. We distinguish between the core \emph{database}, which provides mechanisms for efficiently indexing and searching over dynamic data, and the \emph{DBMS}, which is software that accesses data stored in a database. A database's primary purpose is efficient storage and retrieval of data. DBMSs perform many other functions as well: enforcing data access policies, defining data structures, providing external applications with strong transaction guarantees, serving as building blocks in complex applications (such as visualization and data presentation), replicating data, integrating disparate data sources, and backing up important sources. Recently introduced DBMSs also perform analytics on stored data. We concentrate on the database's core functions of data insertion, indexing, and search.

As the scale, value, and centralization of data increase, so too do security and privacy concerns. 
There is demonstrated risk that the data stored in databases will be compromised.
Nation-state actors target other governments' systems, corporate repositories, and individual data for espionage and competitive advantages~\cite{apt1}.   Criminal groups create and use underground markets to buy and sell stolen personal information~\cite{Motoyama}.  Devastating attacks occur against government~\cite{CyberAttacksOPM} and commercial~\cite{CyberAttacks} targets.

\emph{Protected database search} technology cryptographically separates the roles of providing, administering, and accessing data.  It reduces the risks of a data breach, since the server(s) hosting the database can no longer access data contents.  
 Whereas most traditional databases require the server to be able to read all data contents in order to perform fast search and retrieval, protected search technology uses cryptographic techniques on data that is encrypted or otherwise encoded, so that the server can quickly answer queries without being able to read the plaintext data. 



\subsection{Protected Search Systems Today}

Protected database search has reached an inflection point in maturity. In 2000, Song, Wagner, and Perrig provided the first scheme with communication proportional to the description of the query and the server performing (roughly) a linear scan of the encrypted database~\cite{SP:SonWagPer00}. Building on this, the field quickly moved from theoretical interest to the design and implementation of working systems.

Protected database search solutions encompass a wide variety of cryptographic techniques, including property-preserving encryption \cite{EC:PanRou12}, searchable symmetric encryption \cite{CCS:CGKO06}, private information retrieval by keyword~\cite{EPRINT:ChoGilNao98}, and techniques from oblivious random access memory (ORAM) \cite{STOC:Goldreich87}.  Like the cryptographic elements used in their construction, protected search systems provide provable security based on the hardness of certain computational problems.  
Provable security comes with several other benefits: a rigorous definition of security, a thorough description of protocols, and a clear statement of assumptions.

Many of these systems have been implemented. Protected search implementations have been tested and found to scale moderately well, reporting performance results on datasets of billions of records~\cite{EPRINT:PodBoePop16,CACM:PRZB12,SP:PKVKMC14,SP:FVKKKM15,C:CJJKRS13,CCS:JJKRS13,NDSS:CJJJKR14,ESORICS:FJKNRS15,RSA:IKLO16}.

In the commercial space, a number of
established and startup companies offer products with protected search functionality, including
Bitglass~\cite{bitglass}, Ciphercloud~\cite{ciphercloud}, CipherQuery~\cite{cipherquery}, Crypteron~\cite{crypteron}, IQrypt~\cite{iqrypt}, Kryptnostic~\cite{kryptnostic}, Google's Encrypted BigQuery~\cite{encbigquery}, Microsoft's SQL Server 2016~\cite{encServer2016}, Azure SQL Database~\cite{azureServer2016}, PreVeil~\cite{preveil}, Skyhigh~\cite{skyhigh}, StealthMine~\cite{stealthmine}, and ZeroDB~\cite{zerodb}.  While not all commercial systems have undergone thorough security analysis, their core ideas come from published work.  For this reason, this paper  focuses on systems with a published description.

Designing a protected search system is a balance between security, functionality, performance, and usability. Security descriptions focus on the information that is revealed, or \emph{leaked}, to an attacker with access to the database server. Functionality is primarily characterized by the query types that a protected database can answer. Queries are usually expressed in a standard language such as the structured query language (SQL).  Performance and usability are affected by the database's data structures and indexing mechanisms, as well as required computational and network cost.

There are a wide range of protected database systems that are appropriate for different applications.
With such a range of choices, it is natural to ask: Are there solutions for every database setting? If so, which solution is best?

\subsection{Our Contribution}

The answers to these questions are complex.  Protected search replicates the functionality of some database paradigms, but the unprotected database landscape is diverse and rapidly changing.  Even for database paradigms with mature protected search solutions, making an informed choice requires understanding the tradeoffs.

The goal of this work is twofold: first, to inform protected search designers on the current and future state of database technology, enabling focus on techniques that will be useful in future DBMSs, and second, to help security and database experts understand the tradeoffs between protected search systems so they can make an informed decision about which technology, if any, is most appropriate for their setting.

We accomplish these goals with the following contributions:
\begin{enumerate}
\item A characterization of database search functionality in terms of base and combined queries.  Traditional databases efficiently answer a small number of queries, called a \emph{basis}. Other queries are answered by combining these basis operations~\cite{codd1970relational}.  Protected search systems have implicitly followed this \emph{basis} and \emph{combination} approach.

Although there are many database paradigms, the number of distinct bases of operations is small.  We advocate for explicitly adopting this basis and combination approach.
\item An identification of the bases of current protected search systems and black-box ways to combine basis queries to achieve richer functionality.
We then put protected search in the context of unprotected search by identifying basis functions currently unaddressed by protected search systems.
\item An evaluation of current attacks that exploit \emph{leakage} of various protected search approaches to learn sensitive information. This gives a snapshot of the current security of available base queries.
\item A roadmap and tools for transforming a protected search system into a protected DBMS capable of deployment.  We present an open-source software package developed by our team that aids with performance evaluation; our tool has evaluated protected search at the scale of 10TB of data. We also present preliminary user opinions of protected search.  Lastly, we summarize systems that have made the transition to full systems, and we challenge other designers to think in terms of full DBMS functionality.
\end{enumerate}

\subsection{Organization}
The remainder of this work is organized as follows: \secref{sec:framework} introduces background on databases and protected search systems, \secref{sec:custom} describes protected search base queries and leakage attacks against these queries, \secref{sec:extending func} describes techniques for combining base queries and discusses remaining functionality gaps, \secref{sec:queries-to-systems} shows how to transform from queries to a full system, and \secref{sec:conclusion} concludes.


\section{Overview of Database Systems}
\label{sec:framework}

This section provides background on the databases and protected search systems that we study in this paper. We first describe unprotected database paradigms and their query bases. Next we define the types of users and operations of a database. We then describe the protected search problem, including its security goals and the security imperfections known as leakage that schemes may exhibit. Finally, we give examples of common leakage functions found in the literature.

\subsection{Database Definition and Evolution}
\label{sec:db trends}

A database is a partially-searchable, dynamic data store that is optimized to respond to targeted queries (e.g., those that return less than 5\% of the total data). Database servers respond to queries through a well established API. Databases typically perform search operations in time sublinear in the database size due to the use of parallel architectures or data structures such as binary trees and B-trees.

Several styles of database engines have evolved over the past few decades. Relational or SQL-style databases dominated the database market from the 1970s to the 1990s.  Over the past decade, there has been a focus on databases systems that support many sizes of data management workloads~\cite{stonebraker2005one}. NoSQL and NewSQL have emerged as new database paradigms, gaining traction in the database market~\cite{ullman1982first,stonebraker1988readings}.

\subsubsection{SQL} Relational databases (often called SQL databases) typically provide strong transactional guarantees and have a well known interface. Relational databases are vertically scalable: they achieve better performance through greater hardware resources. SQL databases comply with ACID (Atomicity, Consistency, Isolation, and Durability) requirements~\cite{haerder1983principles}.


\subsubsection{NoSQL} NoSQL (short for ``not only SQL'') databases emerged in the mid 2000s.  NoSQL optimizes the architecture for fast data ingest, flexible data structures, and relaxed transactional guarantees~\cite{chang2008bigtable}. These changes were made to accommodate increasing amounts of unstructured data. NoSQL databases, for the most part, excel at horizontal scaling and when data models closely align with future computation.


\subsubsection{NewSQL} NewSQL systems bring together the scalability of NoSQL databases and the transactional guarantees of relational databases~\cite{pavlos}. Several NewSQL variants are being developed, such as in-memory databases that closely resemble the data models and programming interface of SQL databases, and array data stores that are optimized for numerical data analysis.

\subsubsection{Future Systems} We expect the proliferation of customized engines that are tuned to perform a relatively small set of operations efficiently.  
While these systems will have different characteristics, we believe that each system will efficiently implement a small set of basis operations.  There are several federated or polystore systems being developed~\cite{elmore2015demonstration,gadepally2016bigdawg,halperin2014demonstration}.

The heterogeneous nature of current and future databases demands a variety of protected search systems. While providing such variety is challenging, there are a small number of base operations that can be combined to provide much of the functionality of the aforementioned systems.

\subsection{Query Bases}
To reduce the space of possible queries that must be secured, we borrow an approach from developers of software specifications and mathematical libraries~\cite{van2008science}. In these fields, it is common to determine a core set of low-level kernels and then express other operations using these kernels.  Similarly, many database technologies have a \emph{query basis}: a small set of \emph{base} operations that can be \emph{combined} to provide complex search functionality.  Furthermore, multiple technologies share the same query basis.  In some cases the basis was not explicit in the original design but was formalized in later work.  Apache Accumulo's native API does not have a rigorous mathematical design, but frameworks such as D4M~\cite{kepner2014adjacency,gadepally2015d4m} and Pig~\cite{olston2008pig} used to manipulate data in Accumulo do.

Leveraging an underlying query basis will allow the protected search community to keep pace with new database systems.  
We discuss three bases found in database systems. First, relational algebra forms the backbone of many SQL and NewSQL systems~\cite{pavlos}. Second, associative arrays provide a mathematical basis for SQL, NoSQL, and NewSQL systems~\cite{kepner2016associative}. Third, linear algebraic operations form a basis for some NewSQL databases.  These bases and database paradigms are summarized in \tabref{tab:db classes}.

  \begin{table*}
  \begin{centering}
  \footnotesize
  \begin{tabular}{ p{2.15cm} || p{1.75cm} | p{3.4cm} |  p{3.2cm} | p{3.0cm} | p{1.9cm} }
Query Basis & Technology  & Fundamental characteristics & Strengths & Weaknesses & Examples \\\hline\hline

\textbf{Rel. Algebra} \newline Set Union \newline Set Difference
& \textbf{SQL}~\cite{codd1970relational}: \newline Relational
& Transaction support, \newline ACID guarantees, \newline Table representation of data
& Popular interface, \newline Common data model~\cite{abadi2009data}
& Upfront schema design, \newline Low insert \& query rate
& MySQL~\cite{mysql} \newline Oracle DB~\cite{loney2004oracle} \newline Postgres~\cite{stonebraker1986design} \\ \cline{2-6}

Products/Joins \newline Projection \newline Selection & \textbf{NewSQL}~\cite{pavlos}: \newline Relational
& Use of in-memory, \newline new system arch.  \newline or simplified data model
& Popular interface, \newline Transactional support, \newline ACID guarantees
& Req. expensive hardware, \newline Often only relational data model
&  Spanner~\cite{corbett2013spanner} \newline MemSQL~\cite{shamgunov2014memsql} \newline Spark SQL~\cite{armbrust2015spark}\\ \cline{2-6}

 & \textbf{Federated}~\cite{carey1995towards}
& Relational model, \newline Partitioned/replicated tables
& Transactional support, \newline High performance, \newline ACID guarantees
& Upfront schema design, \newline Often only relational data model
& Garlic~\cite{carey1995towards} \newline  DB2~\cite{ibmdb2} \\ \hline

\textbf{Assoc. Array Alg.} \newline Construct \newline Find \newline Array $(+, \times)$ \newline Element-wise $\times$
& \textbf{NoSQL}~\cite{chang2008bigtable}: \newline Key-value
& Horizontal scalability, \newline Data rep. as key-value pairs, \newline BASE guarantees~\cite{pritchett2008base}
& High insert rates, \newline Cell-level security \newline Flexible schema
& Sacrifice one of the following: consistency, availability, or partition tolerance
& BigTable~\cite{chang2008bigtable} \newline Accumulo~\cite{kepner2014achieving} \newline HBase~\cite{george2011hbase} \newline mongoDB~\cite{mongodb}\\ \hline

\textbf{Linear Algebra } \newline Construct \newline Find \newline Matrix $(+, \times)$
& \textbf{NoSQL}~\cite{webber2012programmatic}: \newline Graph Databases
& Data represented as adjacency \newline or incidence matrix, \newline Horizontal scalability, \newline Graph operation API
& Natural data representation, \newline Amenable to graph algs.
& Performance, \newline Diverse data models, \newline Difficult to optimize
& Neo4j~\cite{webber2012programmatic} \newline System G~\cite{ibmsystemg} \\ \cline{2-6}

Element-wise $\times$ & \textbf{NewSQL}~\cite{brown2010overview}: \newline Array Databases
& ACID guarantees, \newline Data represented as arrays \newline (dense or sparse)
& High performance, \newline Transactional support, \newline Good for scientific comp.
& Data model restrictions, \newline Lack of iterator support
& SciDB~\cite{brown2010overview} \newline TileDB~\cite{li2012scalable} \\ \hline

Multiple bases
& \textbf{Polystore}~\cite{elmore2015demonstration}
& Disparate DBMSs
& High performance, \newline Flexible data stores, \newline Diverse data/programming models
& Requires middleware
& BigDAWG~\cite{elmore2015demonstration} \newline Myria~\cite{halperin2014demonstration} \\ \hline

        \end{tabular}
        \vspace{.05in}
	\caption{Summary of a (not exhaustive) set of popular current and emerging query bases together with their corresponding database technologies.  Characteristics, strengths, weaknesses, and examples refer to the technologies, not the query bases.}
    \label{tab:db classes}
    \end{centering}
  \end{table*}

\subsubsection{Relational Algebra}
Relational algebra, proposed by Codd in 1970 as a model for SQL~\cite{codd1970relational}, consists of the following primitives: set union, set difference, Cartesian product (joins), projection, and selection. Complex queries can typically be generated by composing these operations. Relational algebra and the composability of operations allow a server-side query planner to optimize query execution by rearranging operations to still produce the same result~\cite{smith1975optimizing}.


\subsubsection{Associative Arrays}
\emph{Associative arrays} are a mathematical basis for several styles of database engines~\cite{kepner2016associative}. They provide a mathematical foundation for key-value store NoSQL databases.  Associative array algebra consists of the following base operations: construction, find, associative array addition, associative array element-wise multiplication, and array multiplication~\cite{kepner2014adjacency}.  Associative arrays are built on top of the algebraic concept of a semiring (a ring without an additive inverse).  Addition or multiplication in an associative array can denote any two binary operations from an algebraic semiring. Usually, these two operations are the traditional $\times$ and $+$, but in the min-plus algebra the two operations are $\min$ and $+$ (in the $\max$-plus algebra the two operations are $\max$ and $+$).


\subsubsection{Linear Algebra}

A number of newer NewSQL databases support linear algebraic operations. GraphBLAS is a current standardization effort underway for graph algorithms~\cite{kepner2015graphs}. 
In GraphBLAS, graph data is stored using sparse matrices, and   
the linear algebraic base operations of construction, find, matrix addition, matrix multiplication, and element-wise multiplication are composed to create graph algorithms. Examples of how the GraphBLAS can be applied to popular graph algorithms are given in~\cite{gadepally2015gabb,hutchison2015graphulo}.


\subsection{Database Roles and Operations}

We consider five important database \emph{roles}, analogous to roles in database systems like Microsoft SQL Server 2016 \cite{SQLServer16Roles}:
\begin{itemize}

\item A \textbf{provider}, who provides and modifies the data. 

\item A \textbf{querier}, who wishes to learn things about the data.

\item A \textbf{server}, who handles storage and processing.

\item An \textbf{authorizer}, who specifies data- and query-based rules.

\item An \textbf{enforcer}, who ensures that rules are applied.
\end{itemize}

Databases provide an expressive language for representing permissions, or \emph{rules}. Rules are enforced by authenticating the roles possessed by a valid \emph{user} and granting her the appropriate powers. In general, each user may perform multiple roles, and each role may be performed by multiple users.

While databases offer a wide range of features, we focus on four operations: $\Init$, $\Query$, $\Update$, and $\Rebuild$. These operations are common across the database paradigms described above; we describe them below in the context of protected search.
\begin{itemize}
\item $\Init$: The initialization protocol occurs between the provider (who has data to load) and the server. The server obtains a protected database representing the loaded data.
\item $\Query$: The query protocol occurs between the querier (with a query), the server (with the protected database), the enforcer (with the rules), 
 and possibly the provider. The querier obtains the query results if the rules are satisfied.
\item $\Update$: The update protocol occurs between the provider (with a set of updates) and the server. The server obtains an updated protected database.  Updates include insertions, deletions, and record modifications.
\item $\Rebuild$: The refresh protocol occurs between the provider and the server. The server obtains a new protected database that represents the same data but is designed to achieve better performance and/or security.
\end{itemize}
All systems considered in this work support $\Init$ and $\Query$, but only some systems support \Update and \Rebuild; see Tables \ref{tab:base-query} and \ref{tab:dbms} for details.

\subsection{Protected Database Search Systems}

Informally, a protected search system is a database system that supports the roles and operations defined above, in which each party learns only its intended outputs and nothing else. In particular, a protected search system aims to ensure that the server learns nothing about the data stored in the protected database or about the queries, and the querier learns nothing beyond the query results. These security goals
can be formalized using the \emph{real-ideal} style of cryptographic definition. In this paradigm, one imagines an ideal version of a protected search system, in which a trusted external party performs storage, queries, and modifications correctly and reveals only the intended outputs to each party. The real system is said to be secure if no party can learn more from its real world interactions than it can learn in the ideal system.

We restrict our attention in this work to protected database search systems that provide formally defined security guarantees based upon the strength of their underlying cryptographic primitives.
Some of the commercial systems mentioned in the introduction lack formal security reductions; although they are based on techniques with proofs of security, analysis is required to determine whether differences from those proven techniques affect security.

\subsubsection*{Scenarios}
Only a few existing protected search systems consider the enforcement of rules (i.e., include an authorizer and enforcer). 
Therefore, in this paper we focus primarily on two scenarios: a
\emph{three-party} scenario comprising a provider, a querier, and a server, and a \emph{two-party} scenario in which a single user acts as both the provider and the querier (we denote this combined entity as the \emph{client}).
The latter scenario models a cloud storage outsourcing application in which a client uploads files to the cloud that she can later request. In the two-party setting, the client has the right to know all information in the database so it is only necessary to consider
security against an adversarial server.
In this work, we focus on protected search in the case of a single provider and a single querier; for the more general setting in which multiple users can perform a single role, see \secref{sec:queries-to-systems} and \cite{Bosch14}.

We stress that a secure search system for one scenario does not automatically extend to another scenario. Additionally, despite the limited attention in the literature thus far, we believe that the authorizer and enforcer roles are an important aspect of the continued maturation of protected search systems; see \secref{ssec:rule enforcement} for additional discussion.

\subsubsection*{Threats}
\label{sub:threat-model}
There are two types of entities that may pose security threats to a database: a valid user known as an \emph{insider} who performs one or more of the roles, and an \emph{outsider} who can monitor and potentially alter network communications between valid users. We distinguish between adversaries that are \emph{semi-honest} (or \emph{honest-but-curious}), meaning they follow the prescribed protocols but may passively attempt to learn additional information from the messages they observe, and those that are \emph{malicious}, meaning they are actively willing to perform any action necessary to learn additional information or influence the execution of the system. An outsider adversary (even a malicious one) can be thwarted using secure point-to-point channels. Furthermore, we distinguish between adversaries that \emph{persist} for the lifetime of the database and those that obtain a \emph{snapshot} at a single point in time \cite{CCS:GMNRS16}. The bulk of active research in protected search technology considers semi-honest security against a persistent insider adversary.


\subsubsection*{Performance and Leakage}
While unprotected databases are often I/O bound, protected systems may be compute or network bound. We can measure the performance of a protected operation by calculating the computational overhead and the additional network use (in both the number of messages and the total amount of data transmitted). The type of cryptographic operations matters as well: whenever possible, slower public-key operations should be avoided or minimized in favor of faster symmetric-key primitives.

In order to improve performance, many protected search systems reveal or \emph{leak} information during some or all operations.
Leakage should be thought of as an imperfection of the scheme. The real-ideal security definition is parameterized by the system's specific \emph{leakage profile}, which comprises a sequence of functions that formally describe all information that is revealed to each party beyond the intended output. A security proof demonstrates that the claimed leakage profile is an upper bound on what is actually revealed to an adversary.  Protected search systems' security is primarily distinguished by their leakage profile; our security discussion focuses on leakage.

While leakage profiles are comprehensive, it is often difficult to interpret them and to assess their impact on the security of a particular application (see \secref{sec:leakage-attack}).  To help with this task, the next section identifies common types of leakage.


\subsection{Common Leakage Profiles}
\label{sec:common leakage}

This section provides a vocabulary (partially based on Kamara~\cite{seny_bertinoro}) to describe common features of leakage systematically. While the exact descriptions of leakage profiles are often complex, their essence can mostly be derived from four characteristics: the objects that leak, the type of information leaked about them, which operation leaks, and the party that learns the leakage.

The following types of objects within a protected search system are vulnerable to leakage.
\begin{enumerate}
\item \label{enum:1} Data items, and  any indexing data structures.
\item Queries.
\item Records returned in response to queries, or other relationships between the data items and the queries (e.g., records that partially match a conjunction query).
\item \label{enum:4} Access control rules and the results of their application.
\end{enumerate}
%

Next, we categorize the information leaked from each object. The leakage may occur independently for each query or response, or it may depend upon the prior history of queries and responses. For complex queries like Booleans, leakage may also depend on the connections between the clauses of a query.
While the details of leakage may depend on the specific data structures used for representing and querying the data, we list five general categories of information that may be leaked from objects, ranked from the least to most damaging. We use this ranking throughout our discussion of base queries.
\begin{itemize}

\item[\none] Structure: properties of an object only concealable via padding, such as the length of a string, the cardinality of a set, or the circuit or tree representation of an object.

\item[\oneQuarter] Identifiers: pointers to objects so that their past/future accesses are identifiable.

\item[\half] Predicates: identifiers plus additional information on objects. Examples include ``matches the intersection of 2 clauses within a query'' and ``within a common (known) range.''

\item[\threeQuarter] Equalities: which objects have the same value.

\item[\full] Order (or more): numerical or lexicographic ordering of objects, or perhaps even partial plaintext data.

\end{itemize}

Each of the four database operations may leak information.
During $\Init$, the server may receive leakage about the initial data items.
Every party may receive leakage during a
$\Query$: the querier may learn about the rules and the current data items; the server may learn about the query, the rules, and the current data items; the provider may learn about the query and rules; and the enforcer may learn about the query and current data items.
During $\Update$, the server may receive leakage about the prior and new data records.
During a $\Rebuild$, the server may receive leakage about the current data items.

In a two-party protected search system without \Update or rules it suffices to describe the leakage to the server during \Init and \Query.
In this setting, common components of leakage profiles include:
equalities of queries (often called \emph{search patterns});
identifiers of data items returned across multiple queries 
(often called \emph{access patterns}); the circuit topology of a boolean query; and cardinalities and lengths of data items, queries, and query responses.
Dynamic databases must also consider leakage during $\Update$ and $\Rebuild$. Three-party databases with access restrictions must also consider leakage to the
provider and querier about any objects they didn't produce themselves.



\subsection{Comparison with Other Approaches}

We intentionally define protected database search by its objective rather than the techniques used. As we will see in \secref{sec:custom}, many software-based techniques suffice to construct protected database search. Many hardware-based solutions like \cite{bajaj2014trusteddb} are viable and valuable as well; however, they use orthogonal assumptions and techniques to software-only approaches. To maintain a single focus in this SoK, we restrict our attention to software-only approaches.

Within software-only approaches, the cryptographic community has developed several general primitives that address all or part of the protected database search problem.
\begin{itemize}
\item Secure multi-party computation~\cite{FOCS:Yao82b,STOC:BenGolWig88,STOC:GolMicWig87}, fully homomorphic encryption~\cite{STOC:Gentry09,ITCS:BraGenVai12,PKC:GenHalSma12}, and functional encryption~\cite{FOCS:GGHRSW13} hide data while computing queries on it.

\item Private information retrieval \cite{FOCS:CGKS95,STOC:GIKM98,EPRINT:ChoGilNao98} and oblivious random-access memory (ORAM) \cite{STOC:Goldreich87} hide access patterns over the data retrieved. On their own, they typically do not support searches beyond a simple index retrieval; however, several schemes we discuss in the next section use ORAM in their protocols to hide access patterns while performing expressive database queries.

%

\end{itemize}
Protected search techniques in the literature often draw heavily from these primitives, but rarely rely exclusively on one of them in its full generality.
Instead, they tend to use specialized protocols, often with some leakage, with the goal of improving performance.

Another related area of research known as authenticated data structures ensures correctness in the presence of a malicious server but does not provide confidentiality (e.g.,~\cite{RSA:GTTC03,ICICS:PapTam07,ICISC:EteKup13,SP:BBFR15,TCC:ABCHSW12}).  In general, authenticated data structures do not easily compose with protected database search systems.

\section{Base Queries}
\label{sec:custom}
\ben{Mayank: you want to cite monomi+ mylar somewhere}

\mv{Arkady and I are wondering if we're exclusively focused on the `where' clause of a SQL query, or if we want to handle other types like `group by' (aka ranking). Maybe list this as an open question.}

In this section, we identify basis functions that currently exist in protected search.  The section provides
systematic reviews of the different cryptographic approaches used across query types
and an evaluation of known attacks against them. 

Due to length limitations, we focus on the \emph{Pareto frontier} of schemes providing the currently best available combinations of functionality, performance, and security.  This means that we omit any older schemes that have been superseded by later work.
For a historical perspective including such schemes, we refer readers to relevant surveys~\cite{HSSVYY16,Bosch14}.

We categorize the schemes into three high-level approaches.
The \ppe (or \shortppe) approach can be used with an unprotected database server; it merely modifies the provider's data insertions and the querier's requests. However, this backwards compatibility comes at a cost to security. The \sse  (or \shortsse) approach achieves lower leakage at the expense of designing special-purpose protected indices together with customized protocols that enable the querier and server to traverse the indices together.
We highlight a third approach \oram (or \shortoram), which is a subset of \shortsse that provides stronger security by obscuring object identifiers (i.e., hiding repeated retrieval of the same record).



\subsection{Base Query Implementations}
\label{sec:base queries}

\ay{Should we have pointers to row of the table to indicate which schemes in the discussion correspond to which row?  This is useful as we refer to papers by author in the text, but then use the scheme name in the table, making it difficult to figure out which ones match.  Is there a good way to set up an index on table rows, so we dont have to do this manually?}
\emily{There is a lot of information in the table that isn't introduced in the text. I think in the text we should first describe the dimensions of protected search systems we care about (the columns) and then describe the schemes.}
\discuss{SPAR-like technologies, especially performing boolean queries with strong security.}

\mv{Use Ariel's book chapter text for Blind Seer and ESPADA?}
\mv{Later: replace Arkady's eprint references with the published conference versions where they exist.}

Cryptographic protocols have been developed for several classes of base queries.  The most common constructions are for equality, range, and boolean queries (which evaluate boolean expressions over equality and/or range clauses), though additional query types have been developed as well.
Here, we summarize some of the techniques for providing these functionalities, splitting them based on the approach used. 

The text below focuses on the distinct benefits of each base query mechanism; \tabref{tab:base-query} systematizes the common security, performance, and usability dimensions along which each scheme can be measured.
From a security point of view, we list the index approach, threat model (cf.~Section \ref{sub:threat-model}), and the amount of leakage that the server learns about the data items during $\Init$ and $\Query$ (cf.~Section \ref{sec:common leakage}). Performance and usability are described along three dimensions: the \emph{scale} of updates and queries that each scheme has been shown to support, the type and amount of \emph{cryptography} required to support updates and queries, and the \emph{network} latency and bandwidth characteristics.
\mv{Still need to update the prior paragraph}

%

\subsubsection{\shortppe}
Property-preserving encryption~\cite{EC:PanRou12} produces ciphertexts that preserve some property (e.g., equality or order) of the underlying plaintexts. Thus, protected searches (e.g., equality or range queries) can be supported by inserting ciphertexts into a traditional database, without changing the indexing and querying mechanisms. As a result, \shortppe schemes immediately inherit decades of advances and optimizations in database management systems.

\paragraph*{Equality}
Deterministic encryption (DET)~\cite{C:BelBolONe07, CACM:PRZB12} \ay{9} applies a randomized-but-fixed permutation to all messages so equality of ciphertexts implies equality of plaintexts, enabling lookups over encrypted data. All other properties are obscured.
However, deterministic encryption typically reveals equalities between data items to the server even without the querier making any queries.

\paragraph*{Range}
Order-preserving encryption (OPE)~\cite{SIGMOD:AKSX04,EC:BCLO09,C:BolCheONe11}\ay{13} preserves the relative order of the plaintexts, enabling range queries to be performed over ciphertexts. This approach requires no changes to a traditional database, but comes at the cost of quite significant leakage: roughly, in addition to revealing the order of data items, 
it also leaks the upper half of the bits of each message \cite{EC:BCLO09}.
Improving on this, Boldyreva et al.~\cite{C:BolCheONe11} show how to hide message contents until queries are made against the database. Mavroforakis et al.~\cite{DBLP:MCOK15} further strengthen security using fake queries.
Finally, mutable OPE~\cite{SP:PopLiZel13}\ay{14} reveals only the order of ciphertexts at the expense of added interactivity during insertion and query execution. 

Many \shortppe approaches can easily be extended to perform boolean queries and joins by simply combining the results of the equality or range queries over the encrypted data. CryptDB~\cite{CACM:PRZB12} handles these query types using a layered or \emph{onion} approach that only reveals properties of ciphertexts as necessary to process the queries being made. They demonstrate at most 30\% performance overhead over MySQL, though this value can be much smaller depending on the networking and computing characteristics of the environment.

\shortppe approaches have been adopted industrially \cite{DBLP:GrofigHHKKSST14} and deployed in commercial systems~\cite{bitglass,ciphercloud,cipherquery,crypteron,iqrypt,kryptnostic,encbigquery,encServer2016,azureServer2016,preveil,skyhigh,stealthmine,zerodb}. \ben{Mayank: I added all the citations from the intro.  I believe they all do \shortppe things.} However, as we will explain in \secref{sec:leakage-attack} and Table~\ref{tab:attacks}, even the strongest \shortppe schemes reveal substantial information about queries and data to a dedicated attacker.


\subsubsection{\shortsse Inverted Index}
Several works over the past decade support equality searches on single-table databases via a reverse lookup that maps each keyword to a list of identifiers for the database records containing the keyword (e.g., \cite{CCS:CGKO06,AC:ChaKam10}).
Newer works provide additional features and optimizations for such equality searches.  Blind Storage~\cite{SP:NavPraGun14}\ay{3} shows how to do this with low communication and a very simple server, while Sophos~\cite{CCS:Bost16}\ay{4} shows how to achieve a notion of forward security hiding whether new records match older queries (this essentially runs \Rebuild on every \Insert).

OSPIR-OXT~\cite{C:CJJKRS13,CCS:JJKRS13,NDSS:CJJJKR14,ESORICS:FJKNRS15}\ay{11} additionally supports boolean queries: the inverted index finds the set of records matching the first term in a query, and a second index containing a list of (record identifier, keyword) pairs is used to check whether the remaining terms of the query are also satisfied. Cryptographically, the main challenge is to link the two indices obliviously, so that the server only learns the connections between terms in the same query.
Going beyond boolean queries, Kamara and Moataz~\cite{EPRINT:KamMoa16} intelligently combine several inverted indices in order to support the selection, projection, and Cartesian product operations of relational algebra 
with little overhead on top of the underlying inverted index (specifically, only using symmetric cryptography). They do so at the expense of introducing additional leakage. Moataz's Clusion library implements many inverted index-based schemes \cite{ClusionPaper,clusion-github}.

Cash and Tessaro demonstrate that secure inverted indices must necessarily be slower than their insecure counterparts, requiring extra storage space, several non-local read requests, or large overall information transfer \cite{EC:CasTes14}.

\subsubsection{\shortsse Tree Traversal}
Another category of \shortsse schemes uses indices with a tree-based structure.  Here a query is executed (roughly) by traversing the tree and returning the leaf nodes at which the query terminates.
The main cryptographic challenge here is to hide the traversal pattern through the tree, which can depend upon the data and query.

For equality queries, Kamara and Papamanthou~\cite{FC:KamPap13}
show how to do this in a parallelizable manner; with enough parallel processing they can achieve an amortized constant query cost.  Stefanov et al.~\cite{NDSS:StePapShi14}
show how to achieve forward privacy using a similar approach.

The BLIND SEER system~\cite{SP:PKVKMC14,SP:FVKKKM15}\ay{10} supports boolean queries by using an index containing a search tree whose leaves correspond to records in the database, and whose nodes contain (encrypted) Bloom filters storing the set of all keywords contained in their descendants. A Bloom filter is a data structure that allows for efficient set membership queries.  To execute a conjunctive query, the querier and server jointly traverse the tree securely using Yao's garbled circuits \cite{FOCS:Yao86}, a technique from secure two-party computation, following branches whose Bloom filters match all terms in the conjunction.
Chase and Shen \cite{ChaseS15}\ay{20} design a protection method based on suffix trees to enable substring search.

Tree-based indices are also amenable to range searches.
The Arx-RANGE protocol~\cite{EPRINT:BoePodPop16}\ay{16} builds an index for answering range queries without revealing all order relationships to the server. The index stores all encrypted values in a binary tree so range queries can be answered by traversing this tree for the end points.
Using Yao's garbled circuits, the server traverses the index without learning the values it is comparing or the result of the comparison at each stage. Roche et al.'s partial OPE protocol~\cite{CCS:RACY16}\ay{15} provides a different tradeoff between performance and security with a scheme optimized for fast insertion that achieves essentially free insertion and (amortized) constant time search at the expense of leaking a partial order of the plaintexts.
\mv{Are we sure that `Boelter et al' is correct? The author order is different at \url{https://eprint.iacr.org/2016/591}}
\ay{I am citing the range query paper, not the Arx paper here.}

\subsubsection{Other \shortsse Indices}
We briefly mention protected search mechanisms supporting other query types: ranking results of boolean queries \cite{FC:BalOhr15,EPRINT:StrRay15}, calculating the inner product with a fixed vector \cite{TCC:SheShiWat09,ISC:BTHJ12}, and computing the shortest distance on a graph \cite{CCS:MKNK15}.  These schemes mostly work by building encrypted indices out of specialized data structures for performing the specific query computation.  For example, Meng et al.'s GRECS system~\cite{CCS:MKNK15}\ay{18,19} provides several different protocols with different leakage/performance tradeoffs that encrypt a sketch-based (graph) distance oracle to enable secure shortest distance queries.

%

\subsubsection{\shortoram}
This class of protected search schemes aims to hide common results between queries. Oblivious RAM (ORAM) has been a topic of research for twenty years~\cite{goldreich1996software} and the performance of ORAM schemes has progressed steadily. Many of the latest implementations are based on the Path ORAM scheme~\cite{CCS:SDSFRY13}. However, applying ORAM techniques to protected search is still challenging~\cite{EPRINT:Naveed15}.

\shortoram schemes typically hide data identifiers across queries by re-encrypting and moving data around in a data structure (e.g., a tree) stored on the server.
Several equality schemes use the \shortoram approach.  Roche et al.'s vORAM+HIRB scheme~\cite{SP:RocAviCho16}\ay{6} observes that search requires an ORAM capable of storing varying size blocks since different queries may result in different numbers of results.  They design an efficient variable-size ORAM (vORAM) and combine it with a history independent data structure to build a keyword search scheme.  Garg et al.'s TWORAM scheme~\cite{C:GarMohPap16}\ay{7}  focuses on reducing the number of rounds required by an ORAM-type secure search.  They use a garbled RAM-like~\cite{EC:LuOst13} construction to build a two-round ORAM resulting in a four-round search scheme for equality queries.  Moataz and Blass~\cite{EPRINT:MoaBla15}\ay{21} design oblivious versions of suffix arrays and suffix trees to provide an \shortoram scheme for substring queries.
While offering  greater security, these schemes still tend to be slower than the constructions in the other classes.

An alternative approach is to increase the number of parties.  This approach is taken by Faber et al.'s 3PC-ORAM scheme~\cite{AC:FJKW15}\ay{8} and Ishai et al.'s shared-input shared-output symmetric private information retrieval (SisoSPIR) protocol~\cite{RSA:IKLO16}\ay{17} to support range queries.  3PC-ORAM shows how by adding a second non-colluding server, one can build an ORAM scheme that is much simpler than previous constructions.  SisoSPIR uses a distributed protocol between a client and two non-colluding servers to traverse a (per-field) B-tree in a way that neither server learns anything about which records are accessed.  
By deviating from the standard ORAM paradigm, these schemes are able to approach the efficiency typically achieved by \sse schemes that do not hide access patterns.

%
%
%
%
%
%

\subsubsection{Supporting Updates}
Another important aspect of secure search schemes is whether they support $\Update$. While update functionality is critical for many database applications, it is not supported by many protected search schemes in the \shortsse and \shortoram categories.  Those that support updates do so in one of two ways.  For ease of presentation, consider a newly inserted record.  In most \shortppe schemes the new value is immediately inserted into the database index, allowing for queries to efficiently return this value immediately after insertion.  In many \shortsse schemes, e.g.,~\cite{SP:PKVKMC14}, new values are inserted into a side index on which a less efficient (typically, linear time) search can be used.  Periodically performing \Rebuild incorporates this side index into the main index; however, due to the cost of \Rebuild it is not possible to do this very frequently.  Thus, depending on the frequency and size of updates, update capability may be a limiting functionality of protected search.  In particular, a major open question is to build protected search capable of supporting the very high ingest rates typical of NoSQL databases.  We return to this open problem in \secref{sec:queries-to-systems}.  Roche et al.~\cite{CCS:RACY16} take a step in this direction with a \shortsse scheme for range queries capable of supporting very high insert rates.

\tabref{tab:base-query} systematizes the protected search techniques discussed in this section along with some basic information about the (admittedly nuanced) leakage profiles that they have been proven to meet. There are several correlations between columns of the table; some of these connections reveal fundamental privacy-performance tradeoffs whereas others simply reflect the current state of the art. To provide one example in the latter category: most \shortppe systems leak information at ingestion, whereas most \shortsse only leak information after queries have been made against the system.
The recent Arx-EQ \cite{EPRINT:PodBoePop16}
bucks this trend by requiring the client to remember the frequency of each keyword.

\ay{We need to differentiate between two types of updates.  Ones that build a side-index and periodically reingests (all custom index schemes do this).  And, the other that automatically indexes updated records (as PPE does).  ORAM does actually modify underlying data structure.}



\subsection{Leakage Inference Attacks}
\label{sec:leakage-attack}

In this subsection and \tabref{tab:attacks}, we summarize \emph{leakage inference attacks} that can exploit the leakage revealed by a protected search system in order to recover some information about sensitive data or queries. Hence, this section details the real-world \emph{impact} of the leakage bounds and threat models depicted in \tabref{tab:base-query}. The two tables are connected via a \textsc{join} on the ``$\server$ leakage'' columns: a protected search scheme is affected by an attack if the scheme's leakage to the server is at least as large as the attack's required minimum leakage.

We stress that leakage inference is a new and rapidly evolving field. As a consequence, the attacks in \tabref{tab:attacks} only cover a subset of leakage profiles included in \tabref{tab:base-query}. Additionally, this section merely provides lower bounds on the impact of leakage because attacks only improve over time.

We start by introducing the different dimensions that characterize attack requirements and efficacy. Then, we sketch a couple representative attacks from the literature. Finally, we describe how the provider and querier should use these attacks to inform their choice of a search system that adequately protects their interests.

\label{sec:leakage_section}
\subsubsection{Attack Requirements}
\label{sec:leakage_attack_model}
We classify attacks along four dimensions: attacker goal, required leakage, attacker model, and prior knowledge. The attacker is the server in all of the attacks we consider, except for the Communication Volume Attack of~\cite{KellarisKNO2016}, which can be executed by a network observer who knows the size of the dataset. We expect future research on attacks using leakage available to other insiders.

\paragraph{Attacker Goal} Current attacks try to recover either a set of queries asked by the querier (\emph{query recovery}) or the data being stored at the server (\emph{data recovery}).

\paragraph{Required Leakage} This is the leakage function that must be available to the attacker.  We focus on the common leakage functions on the dataset and responses identified in Section \ref{sec:common leakage}.  Examples include the cardinality of a response set, the ordering of records in the database, and identifiers of the returned records.  Some attacks require leakage on the entire dataset while others only require leakage on query responses.  

\paragraph{Attacker Model} Current inference attacks assume one of two attacker models. The first is a \emph{semi-honest} attacker as discussed in \secref{sub:threat-model}.  The second is an attacker capable of \emph{data injection}: it can create specially crafted records and have the provider insert them into the database. Note that this capability falls outside the usual malicious model for the server. The attacker's ability to perform data injection depends on the use case. For example, if a server can send an email to a user that automatically updates the protected database, this model is reasonable. On the other hand, it might be harder to insert an arbitrary record into a database of hospital medical records.


\paragraph{Attacker Prior Knowledge} All current attacks assume some prior knowledge, which is usually information about the stored data but may include information about the queries made.  
Attack success is judged by the ability to learn information beyond the prior knowledge.  The following types of prior knowledge (ordered from most to least information) help to execute attacks.
\begin{itemize}
\item[\full] Contents of full dataset:  the data items contained in the database. The only possible attacker goal in this case is query recovery.
\item[\threeQuarter]  Contents of a subset of dataset: a set of items contained in the database.  Both attacker goals are interesting in this case. \item[\half] Distributional knowledge of dataset: information about the probability distribution from which database entries are drawn.  For example, this could include knowledge of the frequency of first names in English-speaking countries.  This type of knowledge can be gained by examining correlated datasets.
\item[\oneQuarter] Distributional knowledge of queries: information about the probability distribution from which queries are drawn.  As above, this might be knowledge that names will be queried according to their frequency in the overall population.
\item[\none] Keyword universe: knowledge of the possible values for each field.
\end{itemize}

Naturally, attacks that require full knowledge of the data are more effective; the reasonableness of this assumption should be evaluated for each use case.

\subsubsection{Attack Efficacy}

We evaluate attack efficacy qualitatively in terms of three metrics:
1) the runtime of the attack, including time required to create any inserted records; 2) the sensitivity of the recovery rate to the amount of prior knowledge; and 3) the keyword universe size attacked.
Note that the strength of an attack is strongly application-dependent; an attack that is devastating on one dataset may be completely ineffective on another dataset.

\tabref{tab:attacks} characterizes currently known attacks based upon their requirements and efficacy. All of the attacks described in the table only require modest computing resources.



\mv{Maybe add update performance?}
\begin{table*}
\centering
\begin{tabular}{l|ll|ccc|cc|ccc|ccc|cc|l}

&&&
\multicolumn{3}{c|}{Threats}&
\multicolumn{2}{c|}{{\server} leakage}&
\multicolumn{3}{c|}{Scale}&
\multicolumn{3}{c|}{Crypto}&
\multicolumn{2}{c|}{Network}&
\\

\rot{Query type}&Scheme (References)&
Approach&\rot{\# of parties}&
\rot{Adversarial \querier}&\rot{Adversarial \server}&
\rot{Init}&\rot{Query}&
\rot{Updatable?}&\rot{Implemented?}&\rot{Scale tested}&
\rot{Crypto type} & \rot{Insert: \# ops} & \rot{Query: \# ops} & \rot{\# round trips} & \rot{Data sent}&
Unique feature\\
\hline


\rots{8}{Equality}

&Arx-EQ~\cite{EPRINT:PodBoePop16}&\shortppe&2&
---&\half&
\none&\oneQuarter&
\full&\yes&\half&
\full&\full&\full& 
\fullPie&\fullPie&
legacy compliant
\\

&Kamara-Papamanthou \cite{FC:KamPap13}&\shortsse&2& 
---&\half& 
\none&\oneQuarter& 
\full&---&---&
\full&\none&\none&
\full&\full&
parallelizable
\\

&Blind Storage~\cite{SP:NavPraGun14}&\shortsse&2&
---&\half&
\none&\oneQuarter&
\full&\yes&\half&
\full&\half&\full&
\half&\full&
low $\server$ work
\\


&Sophos ($\Sigma$o$\phi$o$\varsigma$)\cite{CCS:Bost16}&\shortsse&2&
---&\half& 
\none&\oneQuarter&
\full&\yes&\half&
\none&\half&\full&
\full&\full&
$\Rebuild$ w/ $\Insert$
\\

&Stefanov et al \cite{NDSS:StePapShi14}&\shortsse&2&
---&\half& 
\none&\oneQuarter&
\full&\yes&\half&
\full&\none&\none&
\full&\full&
$\Rebuild$ w/ $\Insert$
\\

&vORAM+HIRB~\cite{SP:RocAviCho16}&\shortoram&2&
---&\half&
\none&\none&
\full&\yes&\full&
\full&\none&\none& 
\nonePie&\oneQuarter&
history independ. 
\\

&TWORAM~\cite{C:GarMohPap16}&\shortoram&2&
---&\half&
\none&\none&
\full&---&---&
\half&\none&\none&
\half&\oneQuarter&
const round
\\

&3PC-ORAM \cite{AC:FJKW15}&\shortoram&3&
\half&\half&
\none&\none&
\full&\yes&\oneQuarter&
\full&\none&\none&
\none&\oneQuarter&
dual \server
\\

\hline

%
%

\rots{4}{Boolean}&
DET~\cite{C:BelBolONe07, CACM:PRZB12}&\shortppe&2&
---&\half&
\threeQuarter&\threeQuarter&
\full&\yes&\full&
\full&\half&\half& 
\fullPie&\fullPie&
supports JOINs
\\

&BLIND SEER~\cite{SP:PKVKMC14,SP:FVKKKM15}&\shortsse&3&
\full&\full&
\none&\half&
\half&\yes&\full&
\half&\half&\none& 
\nonePie&\oneQuarter&
hide field, $r_i$'s
\\

&OSPIR-OXT~\cite{C:CJJKRS13,CCS:JJKRS13,NDSS:CJJJKR14,ESORICS:FJKNRS15,clusion-github}&\shortsse&3&
\full&\half&
\none&\half&
\full&\yes&\full&
\half&\half&\half& 
\threeQuarter&\fullPie&
excels w/ small $r_1$
\\

&Kamara-Moataz~\cite{EPRINT:KamMoa16}&\shortsse&2&
---&\half&
\none&\half&
\none&---&---&
\half&\half&\none& 
\full&\oneQuarter& 
relational SPC
\\

\hline

\rots{5}{Range}
&OPE~\cite{SIGMOD:AKSX04,EC:BCLO09,C:BolCheONe11}&\shortppe&2&
---&\half&
\full&\full&
\full&\yes&\full&
\full&\full&\full& 
\fullPie&\fullPie&
leak some content
\\

&Mutable OPE~\cite{SP:PopLiZel13}&\shortppe&2&
---&\half&
\full&\full&
\full&\yes&\half&
\full&\none&\none& 
\nonePie&\half&
interactive
\\


&Partial OPE~\cite{CCS:RACY16}&\shortsse&2&
---&\half&
\none&\full&
\full&\yes&\half&
\full&\full&\full& 
\half&\fullPie&
fast insertions
\\



&Arx-RANGE~\cite{EPRINT:BoePodPop16}&\shortsse&2&
---&\half&
\none&\half&
\full&\yes&\half&
\half&\none&\none& 
\fullPie&\nonePie&
non-interactive
\\

&SisoSPIR~\cite{RSA:IKLO16}&\shortoram&3&
\half&\half&
\none&\none&
\none&\yes&\full&
\full&\full&\full&
\none&\oneQuarter&
split, non-colluding \server
\\
\hline

%
%
%
%

\rots{4}{Other}
&GraphEnc$_1$~\cite{CCS:MKNK15}&\shortsse&2&
---&\maybeCircle&
\half&\oneQuarter&
\none&\yes&\half&
\full&\full&\full&
\full&\oneQuarter&
approx.~graph dist.
\\

&GraphEnc$_3$~\cite{CCS:MKNK15}&\shortsse&2&
---&\half&
\half&\half&
\none&\yes&\half&
\none&\full&\full&
\full&\full&
approx.~graph dist.
\\

&Chase-Shen \cite{ChaseS15,Cocoon}&\shortsse&2&
---&\full&
\none&\half&
\none&\yes&\oneQuarter&
\full&\full&\full&
\half&\full&
substring search
\\

&Moataz-Blass \cite{EPRINT:MoaBla15}&\shortoram&2&
---&\half&
\none&\none&
\full&\yes&\full&
\full&\none&\none&
\none&\oneQuarter&
substring search
\\

\hline

\end{tabular}
\vspace{.05in}
\caption[Summary of base queries]{\scriptsize Summary of the security, performance, and usability of base queries. \querier and \server denote the querier and the server, respectively.
We presume that the adversary knows the database size $d$ and the length of each record.
For systems that either do not support insert or use a side index, the insert cost is the amortized cost of adding a single record during \Init.
Legends for each column follow. In all columns except ``Init/Query leakage,'' bubbles that are more filled in represent properties that are better for the scheme.

\vspace{.1in}
\begin{tabular}{p{3.2cm}  p{3.2cm} p{2.4cm} p{3.2cm} p{3.2cm}}
Scale Tested & Updatable  & Threats & Data sent &Init/Query leakage\\
 \multirow{3}{*}{\begin{tabular}[c]{@{}l@{}}\fullPie -- Billions\\					
								\halfPie -- millions\\
								\oneQuarter -- thousands
\end{tabular}} &
\multirow{3}{*}{\begin{tabular}[c]{@{}l@{}}\fullPie -- insert in main index\\
								\halfPie -- build side index\\
								\nonePie -- not supported
\end{tabular}} &
\multirow{2}{*}{\begin{tabular}[c]{@{}l@{}}\fullPie -- malicious\\
								\halfPie -- semi-honest
\end{tabular}} &
 \multirow{5}{*}{\begin{tabular}[c]{@{}l@{}}(beyond results)\\\fullPie -- constant\\					
								\half -- additive polylog($d$) \\
								\oneQuarter -- mult. polylog($d)$\\
								\none -- even more
\end{tabular}} &
\multirow{6}{*}{\begin{tabular}[c]{@{}l@{}}	(see \secref{sec:common leakage})\\
								\fullPie -- order/contents\\
								\threeQuarter -- equality\\
								\half -- predicate\\
								\oneQuarter -- identifier\\
								\none -- structure
\end{tabular}} \\\\\\\\
Type of Crypto & Crypto Ops per Record & Round Trips &  &\\
 \multirow{4}{*}{\begin{tabular}[c]{@{}l@{}}
 								\fullPie -- symmetric\\					
								\halfPie -- batched or pre-\\
								computed public-key\\
								\none -- public-key
\end{tabular}} &
 \multirow{3}{*}{\begin{tabular}[c]{@{}l@{}}\fullPie -- constant\\					
								\halfPie -- \# keywords\\
								\none -- logarithmic
\end{tabular}} &
 \multirow{4}{*}{\begin{tabular}[c]{@{}l@{}}\fullPie -- 1\\					
								\threeQuarter -- 2\\
								\half -- constant\\
								\none -- logarithmic
\end{tabular}} &
 \\\\\\\\
\end{tabular}

\ay{Check whether Stealth paper builds upon any of the equality-based ORAMs, such as 3PC-ORAM. Specifically, see whether Stealth cites the other papers.}
}

\label{tab:base-query}

\begin{tabular}{p{1.6cm}| p{.4cm} |p{.6cm}|p{1cm}|p{1.25cm}|p{.9cm}|p{1.3cm}|p{1.2cm}|  p{4.5cm}}
 &  \multicolumn{2}{p{1.2cm} |}{Required $S$ leakage}&  \multicolumn{2}{p{1.9cm} |}{Required attack  conditions}  &  \multicolumn{3}{c|}{Attack efficacy}&\\
Attacker goal& Init&Query&Ability to inject data & Prior knowledge & Runtime & Sensitivity to prior knowledge & Keyword universe tested & Attack name\\
\hline\hline
\multirow{6}{*}{\rotatebox[origin=c]{50}{Query Recovery}}
 & \nonePie & \nonePie  & --- & \oneQuarter  & \fullPie & ? & \nonePie & Communication Volume Attack \cite{KellarisKNO2016} \\ \cline{2-8}
  & \nonePie & \oneQuarter& \yes  & \nonePie & \nonePie & \nonePie & \nonePie & Binary Search Attack \cite{USS:ZhaKatPap16} \\ \cline{2-8}
  & \nonePie & \oneQuarter & --- & \oneQuarter &  \fullPie & ? & \nonePie & Access Pattern Attack \cite{KellarisKNO2016} \\ \cline{2-8}
&  \nonePie & \oneQuarter & ---  & \threeQuarter & \halfPie & \fullPie & \fullPie & Partially Known  Documents \cite{CashGPR15}\\ \cline{2-8}
 & \nonePie &\oneQuarter & \yes & \threeQuarter & \halfPie & \nonePie & \fullPie & Hierarchical-Search Attack \cite{USS:ZhaKatPap16} \\ \cline{2-8}
 & \nonePie & \oneQuarter & --- & \fullPie & \halfPie & \fullPie & \fullPie & Count Attack \cite{CashGPR15} \\ \cline{2-8} \hline
\multirow{5}{*}{\rotatebox[origin=c]{50}{Data Recovery}} & \nonePie & \oneQuarter & --- & \halfPie & \fullPie & \fullPie & \halfPie & Graph Matching Attack \cite{PouliotW16} \\ \cline{2-8}
  & \threeQuarter  & --- & ---  & \halfPie & \nonePie & ? & \nonePie & Frequency Analysis \cite{NaveedKW15} \\ \cline{2-8}
& \threeQuarter & --- & \yes & \halfPie & \nonePie & ? & \fullPie & Active Attacks \cite{CashGPR15}  \\ \cline{2-8}
 & \threeQuarter & --- &  --- &  \threeQuarter & \nonePie & ? & \fullPie & Known Document Attacks \cite{CashGPR15} \\ \cline{2-8}
   & \full  & --- & --- &  \halfPie & \nonePie & \nonePie & \fullPie & Non-Crossing Attack \cite{grubbs2016leakage} \\ \hline
\end{tabular}
\vspace{.05in}
\caption[Summary of Leakage Inference Attacks]{\scriptsize
Summary of current leakage inference attacks against protected search base queries.  \server is the server and the assumed attacker for all attacks listed. $\server$ leakage symbols have the same meaning as in \tabref{tab:base-query}. Each attack is relevant to schemes in \tabref{tab:base-query} with at least the $S$ leakage specified in this table. Some attacks require the attacker to be able to inject data by having the provider insert it into the database. Legends for the rest of the columns follow. In all columns except ``Keyword universe tested,'' bubbles that are more filled in represent properties that are better for the scheme and worse for the attacker.

\vspace{.05in}
\begin{tabular}{l l l  l  l}
Prior knowledge  & Runtime (in \# of keywords) & Sensitivity to prior knowledge & Keyword universe tested\\
\begin{tabular}[c]{@{}l@{}}\fullPie -- Contents of full dataset\\
								\threeQuarter -- Contents of a subset of dataset\\
								\halfPie -- Distributional knowledge of dataset\\
								\oneQuarter -- Distributional knowledge of queries\\
								\nonePie -- Keyword universe
								\end{tabular}
 & \begin{tabular}[c]{@{}l@{}}\fullPie -- More than quadratic\\
 					    \halfPie -- Quadratic \\
					  \nonePie-- Linear \end{tabular}
& \begin{tabular}[c]{@{}l@{}}\fullPie -- High\\
					  \nonePie-- Low \\
? -- Untested		\end{tabular}		
&  \begin{tabular}[c]{@{}l@{}}\fullPie -- $>1000$\\
 					    \halfPie --  $500$ to $1000$\\
					  \nonePie-- $<500$ \end{tabular}
\end{tabular}
}

\label{tab:attacks}
\end{table*}

\subsubsection{Attack Techniques}

Leakage inference attacks against protected search systems have evolved rapidly over the last few years, with Islam et al.\cite{IslamKK12} in 2012 inspiring many other papers.
Most of the attacks in \tabref{tab:attacks} rely on the following two facts: 1) different keywords are associated with different numbers of records, and 2) most systems reveal keyword identifiers for a record either at rest (e.g., DET~\cite{CACM:PRZB12} reveals during $\Init$ if records share keywords) or when it is returned across multiple queries (e.g., Blind Seer~\cite{SP:PKVKMC14} reveals during $\Query$ which returned records share keywords).
To give intuition for how these attacks work we briefly summarize two entries of \tabref{tab:attacks}.

Cash et al.'s \cite{CashGPR15} Count Attack is a conceptually simple way to exploit this information. Assume the attacker has full knowledge of the database and is trying to learn the query.  The attacker sees how many records are returned in response to a query. If that number is unique it can identify the query.  Furthermore, by identifying the query, the attacker learns that every returned record is associated with that keyword.

For example, suppose the attacker learns the first query was for LastName = `Smith'. Now consider a second query for an unknown first name. The query does not return a unique number of records, so the method above cannot be used.
Suppose that FirstName=`John' and FirstName=`Matthew' both return 1000 records.  The attacker can also check how many records are in common with the previous query.  This creates an additional constraint, for example there may be 100 records with name `John Smith' but only 10 records with name `Matthew Smith'.  By checking record overlap with the previously identified query, the attacker can identify the queried first name.  This attack iteratively identifies queries and uses them as additional constraints to identify unknown queries.

Cash et al.'s attack is fairly simple and performs well if the keyword universe sizes is at most  $5000$.  However, it requires a large portion of the dataset to be known to the attacker. With 80\% of the dataset known to the attacker, Cash et al.~\cite{CashGPR15} yield a 40\% keyword recovery rate.

Zhang et al. \cite{USS:ZhaKatPap16} extend the Count Attack to a malicious adversary setting, allowing a server to inject a set of constructed records. This capability greatly improves keyword recovery. By carefully constructing a small number of these records (e.g., nine records for a universe of 5000 keywords), it is possible to search the keyword universe and identify the keyword.
Although the records are fairly large, the attack extends if the database only allows a limited number of keywords per data record. This attack recovers more keywords than the attack of Cash et al.: 40\% of the data must be leaked to obtain a 40\% keyword recovery rate.

\subsubsection{Discussion}
The provider and querier rely upon protected search to protect themselves against the server, or anyone who compromises the server. Our systemization of attacks shows that they should consider the following four questions before choosing a protected search technique to use.
\begin{itemize}
\item How large is the keyword universe?
\item How much of the dataset or query keyword universe (and corresponding frequency) can the attacker predict? 
\item Can an attacker reasonably insert crafted records?
\item Does the adversary have persistent access to the server, or merely a snapshot of it at a single point in time?
\end{itemize}
Answers to the first three questions depend upon the intended use case. For example, a system with a smaller leakage profile may be necessary in a setting where the keyword universe is small and the attacker has the ability to add records.  A system with a larger leakage profile may suffice in a setting where the keyword universe is very large. 

The fourth question pertains to adversaries who compromise the server. \shortppe schemes tend to leak information about the entire database to the server. Thus, using the terminology of Grubbs et al.~\cite{CCS:GMNRS16}, they are susceptible to an adversary who only gets a \emph{snapshot} of the database at some point in time. In contrast, \shortsse schemes tend to reveal information about records only during record retrieval or index modification as part of the querying process, so they require a \emph{persistent} adversary who can observe the evolution of the database state over time. (We note however that many Boolean schemes have additional leakage about data statistics for the entire database.)


In summary, each protected search approach has a distinct leakage profile that results in qualitatively different attacks.
If queries only touch a small portion of the dataset or the adversary only has a snapshot, the impact of leakage from \shortsse systems is less than from \shortppe schemes. If queries regularly return a large fraction of the dataset, this distinction disappears and an \shortoram scheme may be appropriate.
Recently, Kellaris et al.~\cite{KellarisKNO2016} showed an attack on \shortoram schemes, but it requires significantly smaller database and keyword universe sizes than attacks against non-\shortoram schemes.

\mv{Note for later: if we ever expand upon this paper, we can explain why the distinction between snapshot and persistent matters in the protected DB setting even though they're somewhat equivalent in the unprotected setting since databases store logs of transactions. There's a ``kill chain" that requires much more work in the protected space: snapshot access to the machine that stores the database $\to$ access to the logs (which may be on a different machine) $\to$ capacity to read the contents of the logs (cf.~Ben and Sophia's auditing paper) $\to$ persistent ability to view the evolution of the data structure.}

\textbf{Open Problems:} The area of leakage attacks against protected search is expanding.  Published attacks consider attackers who insert specially crafted data records but have not considered an attacker who may issue crafted queries.
Furthermore, all prior attacks have considered the leakage profile of the server.  Future attacks should consider the implications of leakage to the querier and provider.  Current attacks have targeted Equality and Range queries; we encourage the study of leakage attacks on other query types such as Boolean queries.

On the reverse side, it is important to understand what these leakage attacks mean in real-world application scenarios.  Specifically, is it possible to identify appropriate real-world use-cases where the known leakage attacks do not disclose too much information?  Understanding this will enable solutions that better span the security, performance, and functionality tradeoff space.

Lastly, on the defensive side we encourage designers to implement \Rebuild mechanisms.   \Rebuild mechanisms have only been implemented for Equality systems.


\section{Extending Functionality}
\label{sec:extending func}


\subsection{Query Composition}
\label{sec:query comp}\newcounter{numcombiners}


We now describe techniques to combine the base queries described in Section~\ref{sec:custom} (equality, Boolean, and range queries) to obtain richer queries. We restrict our attention to techniques that are black box (i.e., they do not depend on the implementation of the base queries).

As a general principle, schemes that support a given query type by composing base queries tend to have more leakage than schemes that natively support the same query type as a base query. However, using query composition, a scheme that supports the necessary base queries can be extended straightforwardly to support multiple query types, whereas supporting those all as base queries requires significant effort. Thus, we see value in advancing both base and composed queries.


\tabref{tab:composed queries} summarizes the techniques we describe below. In the table and the text, we cite the first work proposing each approach, though we note that several ideas appear to have been developed independently and concurrently.    We defer the description of string queries (substrings and wildcards) to \apref{sec:other combiners}.

  \begin{table*}
  \begin{centering}

    \begin{tabular}{p{3.7cm}  | p{3cm}   | c |   p{6.4cm} | p{.7cm}}
	    Composed Query & Base Query Calls   & Additional Storage & Leakage & Work\\\hline\hline

	   1. Equality (EQ) & 1 range & none &  Same as range & --- \\\hline
	   2. Disjunction (OR) of $k$ EQs (or ranges) & $k$ EQs (or ranges) & none & Identifiers of records matching each clause, if EQ leaks $\geq \oneQuarter$ & ---  \\\hline
	   3. Conjunction (AND) of $k$ EQs  & 1 EQ & $ {\beta \choose k}$ & Same as EQ &  ---\\\hline
	   4. Stemming & 1 EQ   & $1$ &   Identifiers of records sharing stem, if EQ leaks $\geq \oneQuarter$  & ---\\\hline
       5. Proximity & 1 EQ & $\ell$ &  Identifiers of neighbor pairs, if EQ leaks $\geq \oneQuarter$ & \cite{FSE:BolChe14}  \\\hline
       6. Range w/ small domain & $(2 + r)$ EQs &  $1$  & No leakage if refresh between queries &  \cite{CrescenzoG15} \\\hline 
	   7. Range & OR of $(2 \log m)$ EQs & $\log m$  & Distributional info, if EQ leaks $\geq \oneQuarter$ & \cite{SP:PKVKMC14}\\\hline 
	   8. Negation & AND of 2 ranges & 1 & Same as OR of ranges & \cite{SP:PKVKMC14} \\\hline
	    9. Substring ($\rho = \kappa$) 
& 1 EQ  & $\alpha-\kappa+1$ & Identifiers of records sharing $\kappa$-grams, if EQ leaks $\geq \oneQuarter$ & \cite{RSA:IKLO16}\\\hline
	    10. Substring ($\rho \le \kappa$) & 1 range  &$\alpha-\kappa + 1$ & Same as range, on $\kappa$-grams & \cite{RSA:IKLO16} \\\hline
	    11. Anchored Substring ($\rho \ge \kappa$) & AND of $(\rho - \kappa + 1)$ EQs & $\alpha - \kappa + 1 $ & If EQ leaks $\geq \oneQuarter$, rec. ids. w/ $\kappa$-grams in same positions; if AND leaks \# clauses, $\rho$ & \cite{C:CJJKRS13} \\\hline
	    12. Substring & OR of $(\alpha - \kappa + 1)$ ANDs of $(\rho - \kappa + 1)$ EQs &    $\alpha - \kappa + 1 $  &  If EQ leaks $\geq \oneQuarter$, rec. ids. w/ $\kappa$-grams in same positions; if AND leaks \# clauses, $\rho$  & \cite{C:CJJKRS13}\\\hline
	    13. Anchored Wildcard & AND of $(\rho - \kappa + 1)$ EQs &   $\alpha- \kappa + 1$ & If EQ leaks $\geq \oneQuarter$, rec. ids. w/ $\kappa$-grams in same positions; if AND leaks \# clauses, $\rho$ & \cite{C:CJJKRS13} \\\hline
	    14. Wildcard & OR of $(\alpha - \kappa + 1)$ ANDs of $(\rho - \kappa + 1)$ EQs & $\alpha -\kappa + 1 $ & If EQ leaks $\geq \oneQuarter$, rec. ids. w/ $\kappa$-grams in same positions; if AND leaks \# clauses, $\rho$ & \cite{C:CJJKRS13}\\\hline
    \end{tabular}
\vspace{.05in}
    \caption[Summary of Query Combiners]{
    Summary of query combiners using equality (EQ), conjunction (AND), disjunction (OR), and range base query types. Storage is given as additional storage beyond that required for the base equality or range queries, as a multiplicative factor over the base storage. Composed query leakage depends on the leakage of the base queries used; the table gives the composed query leakage if the base equality scheme leaks identities. ``Anchored'' refers to a search that occurs at either the beginning or the end of a string.

   \vspace{.05in}
\begin{tabular}{p{5cm}   p{5cm} p{5cm} }
Boolean Notation & Proximity, Range Notation & String Notation\\
$k$ = \# of clauses in Boolean & $\ell$ =  Max \# of neighbors of a record &  $\kappa$ = Length of grams\\
$\beta$ = Max \# of keywords per record & $m$ = Size of domain & $\rho$ = Length of query string\\
& $r$ = \# query results &  $\alpha$ = Max length of data string (padded if necessary)
\end{tabular}
}
    \label{tab:composed queries}
    \end{centering}
  \end{table*}

\subsubsection{Equality using range}\stepcounter{numcombiners}
Equality queries can be supported using a range query scheme. To obtain the records equal to $a$, the querier performs a range query for the range $[a, a]$.

\subsubsection{Disjunction of equalities/ranges using equality/range} \label{par:or-from-eq}\stepcounter{numcombiners}
Disjunctions of equalities or ranges can be supported using an equality or a range scheme, respectively. To obtain the records that equal any of a set of $k$ keywords $w_1, \ldots, w_k$, the querier can perform an equality query for each keyword $w_i$ and combine the results. Similarly, to retrieve all records that are in any of $k$ ranges, the querier can perform a range query for each range and combine the results. This approach reveals to the server the leakage associated with each equality or range query, e.g., the exact or approximate number of records matching each clause (not just the number of records matching the disjunction overall).
\subsubsection{Conjunction of equalities using equality}\stepcounter{numcombiners}
Conjunctions of equalities can be supported using an equality scheme. To supporting querying for records that match all of the keywords $w_1, \ldots, w_k$, one builds an equality scheme containing $k$-tuples of keywords.
The querier then performs an equality search on the $k$-tuple representing her query to retrieve the records that contain all of those keywords.  The storage for this approach grows exponentially with $k$ but is viable for targeted keyword combinations or a small number of fields.

\subsubsection{Stemming using equality}\stepcounter{numcombiners}
Stemming reduces words to their root form; stemming queries allow matching on word variations. For example, a stemming query for `run' will also return results for `ran' and `running'.
The Porter stemmer is a widely used algorithm~\cite{porter1980algorithm,willett2006porter}.  Stemming can be supported easily by using the stemmed version of keywords at both initialization and query time, and thus performing the match using
a single equality query.

\subsubsection{Proximity using equality}\stepcounter{numcombiners}
Proximity queries find values that are `close' to the search term. Li et al.~\cite{DBLP:LiWWCRL10} support proximity queries by building an equality scheme associating each neighbor of any record with its set of neighbors in the dataset at initialization; a proximity query is then an equality query, which will return a record if it matches the queried value or is a neighbor of it. Boldyreva and Chenette~\cite{FSE:BolChe14} improve on the security of this scheme by revealing only pairwise neighbor relationships instead of neighbor sets.  They also pad the number of inserted keywords to the maximum number of neighbors.
This solution multiplies storage by the maximum number of neighbors of a record. If disjunctive searches are permitted, one can trade off storage space with the number of terms in the search.

Another approach uses locality-sensitive hashing~\cite{indyk1998approximate,gionis1999similarity}, which preserves closeness by mapping `close' inputs to identical values and `far' inputs to uncorrelated values. Proximity queries can be supported by inserting the output of a locality-sensitive hash as a keyword in an equality scheme. Returning only `close' records requires matching the output of multiple hashes.  Parameters vary widely depending on the notion of closeness.  This approach has been demonstrated for Jaccard distance\cite{DBLP:KuzuIK12} and Hamming distance \cite{DBLP:ParkKLCZ07,DBLP:AdjedjBCK09,DBLP:BringerCK09,DBLP:LiWWCRL10,DBLP:WangRYU12}.

\subsubsection{Small-domain range query using equality \cite{CrescenzoG15}}\stepcounter{numcombiners}
To support range queries on a searchable attribute $A$ with domain $D$, we build two equality-searchable indices. The first index maps each value $a\in D$ to the number of records in the database smaller than $a$ and the number of records larger than $a$. With two equality queries into this index, the querier can learn the location of the lower and upper bounds of a range query. The second index is an ordered list of records sorted by $A$, from which the client reads the relevant subset.

This approach requires blinding factors to prevent the client from learning the positions of the results while still being able to search the second index~\cite{CrescenzoG15}. Also, this approach only works for attributes with small domain, since the first index has size proportional to the domain size.

\subsubsection{Large-domain range using equality and disjunction \cite{CrescenzoG15,SP:PKVKMC14}}\stepcounter{numcombiners}

Range queries can be performed over exponential size domains via range covers, which are a specialization of set covers that effectively pre-compute the results of canonical range queries that would be asked during a binary search of each record.
For instance, consider the domain $D=[0,8)$ with size $m =8$. To insert a record with attribute $A=3$, we insert keywords corresponding to each of the canonical ranges $[0,8)$, $[0,4)$, $[2,4)$, and $[3, 4)$.
Range queries are split into canonical ranges; for instance, the range $[2,5)$ would be split into $[2,4)$ and $[4,5)$.
Combining this technique with disjunctions yields range queries~\cite{SP:PKVKMC14}.

Demertzis et al.~\cite{demertzispractical}
provide a variety of range cover schemes
with different tradeoffs between leakage, storage, and computation. At the extremes, they can support constant storage with query cost linear in the range size, or $m^2$ multiplicative storage with constant-sized keyword queries. They recommend a balanced approach similar to~\cite{CrescenzoG15,SP:PKVKMC14}, although their recommended scheme has false positives.

\subsubsection{Negations using range and disjunction \cite{SP:PKVKMC14}}\stepcounter{numcombiners}

As above consider an ordered domain $D$ with minimum and maximum values $a_{min}$ and $a_{max}$, respectively.  To search for all records not matching $A=a$, compute a disjunction of the queries $[a_{min}, a)$ and $(a, a_{max}]$.

\subsection{The Functionality Gap}
\label{sec:gaps in protected search}
We now review gaps in query functionality based on current protected base and combined queries.    Our discussion is divided among the three query bases from \secref{sec:db trends}.

\paragraph{Relational Algebra}
Cartesian product, which corresponds to the JOIN keyword in SQL, has been demonstrated in \shortppe schemes.  The one \shortsse scheme that supports Cartesian product is the work of Kamara and Moataz~\cite{EPRINT:KamMoa16}, but their scheme does not support updates.

The JOIN keyword makes a system \emph{relational}.  Secure JOIN is a crucial capability for protected search systems.  The key challenge is to create a data structure capable of linking different values that reveals no information to any party.  This challenge also arises in Boolean \shortsse systems.  Systems overcome this challenge by placing values that could be linked in a single joint data structure.  It is difficult to scale this approach to the JOIN operation as the columns involved are not known ahead of time (and there are many more possibilities).

\textbf{Open Problem:} Support secure Cartesian product using \shortsse and \shortoram approaches.

\paragraph{Associative Arrays}
The main workhorse of associative arrays is the ability to quickly add and multiply arrays.  \shortppe schemes have shown how to support limited addition through the use of somewhat homomorphic encryption.  There is extensive work on private addition and multiplication using secure computation.  However, this problem has not received substantial attention in the protected search literature.  We see adaptation of (parallelizable) arithmetic techniques into protected search as a key to supporting associative arrays.

\textbf{Open Problem:} Incorporate secure computation into protected search systems to support array $(+, \times)$.

In addition, associative arrays are often constructed for string objects.  In this setting, multiplication and addition are usually replaced with the concatenate function and an application-defined `minimum' function that selects one of the two values.  Finding the minimum is connected to the comparison operation.  The comparison operation has been identified as a core gadget in the secure computation literature~\cite{damgard2008homomorphic,kerschbaum2009performance}.  We encourage adaptation of this gadget to protected search.

\textbf{Open Problem:} Support protected queries to output the minimum of two strings.

\paragraph{Linear Algebra}
The main gap in supporting linear algebra is how to privately multiply two matrices.  This problem is made especially challenging as for different data types the addition ($+$) and multiplication ($\times$) operations may be defined arbitrarily.   Furthermore, linear algebra databases store data as sparse matrices.  Access patterns to a sparse matrix may leak about the contents.  This problem has begun to receive attention in the learning literature \cite{han2008privacy} as matrix multiplication enables many linear classification approaches.  However, current work requires specializing storage to a particular  algorithm, such as shortest path \cite{CCS:MKNK15,CCS:WNLCSS14}.

\textbf{Open Problem:} Support efficient secure matrix multiplication and storage.

\section{From Queries to Database Systems}\label{sec:queries-to-systems}
In addition to search, a DBMS enforces rules, defines data structures, and provides transactional guarantees to an application.
In this section, we highlight important components that are affected by security and need to be addressed to enable a protected search system to become a full DBMS.  We then discuss current protected search systems and their applicability for different DB settings.

\subsection{Controls, Rules and Enforcement}
\label{ssec:rule enforcement}

Classical database security includes a broad set of control measures, including access control, inference control, flow control, and data encryption~\cite{Elmasri2011:FunDatabaseSystems}.

\emph{Access control} assigns a principal such as a user, role, account, or program privileges to interact with objects like tables, records, columns, views, or operations in a given context~\cite{DEPSEC:Bertino2005}.  \emph{Discretionary} access control balances usability with security and is used in most applications. \emph{Mandatory} access control is used where a strict hierarchy is important and available for individuals and data. \emph{Inference control} is used with statistical databases and restricts the ability of a principal to infer a fact about a stored datum from the result returned by an aggregate function such as average or count.  \emph{Flow control} ensures that information in an object does not flow to another object of lesser privilege.  \emph{Data encryption} in classical systems is used for transmitting data from the database back to the client and user.  Some systems also encrypt the data at rest and use fine-grained encryption for access control~\cite{fuchs2012accumulo}.  These techniques are covered in most database textbooks.

A new complementary approach is called \emph{query control}~\cite{spar_baa}.  Query control limits which queries are acceptable, not which objects are visible by a user.  
As an example, a user may be required to specify at least five columns in a query, ensuring the query is sufficiently ``targeted.''
It enables database designers to match legal requirements written in this style.  Query control can be expressed using a \emph{query policy}, which regulates the set of query controls.  

Most current protected search designs do not consider either an \textbf{authorizer} or \textbf{enforcer}.  Integrating this functionality is an important part of maturing protected search and complements the cryptographic protections provided by the basic protocols.


\subsection{Performance Characterization}
\label{sec:eval platform}

Database system adoption depends on response time on the expected set of queries.  Databases are highly tuned, often creating indices on the fly in response to queries.  This makes fair and fast evaluation difficult.
To address this challenge, we developed a performance evaluation platform.  Our platform has been open-sourced with a BSD license (\url{https://github.com/mit-ll/SPARTA}).  Design details can be found in~\cite{HamlinH14,VariaPHHHPRYC15,FCW:VarYakYan15}.
It has been used to test protected search systems at scales of 10TB. Prior works \ifblinded{\cite{redacted5,redacted6,redacted7,redacted8,redacted9}}{\cite{SP:PKVKMC14, SP:FVKKKM15,CCS:JJKRS13,RSA:IKLO16}} report performance numbers generated by our platform.
While the platform has been used to evaluate SQL-style databases it was designed with reusability and extensibility in mind to allow generalization to other types of databases.

Our platform evaluates: 1) integrity of responses and modifications (when occurring individually and while other operations are being performed) and 2) query latency and throughput under a wide variety of conditions.    The system can vary environmental characteristics, the size of the database, query types, how many records will be returned by each query, and query policy.  Each of these factors can be measured independently to create performance cross-sections.

In our experiments, we found protected search response time depends heavily on:
\begin{enumerate}
\item Network capacity, load, and number of records returned by a query.  Protected search systems often have more rounds of communication and network traffic than unprotected databases.
\item The ordering of terms and subclauses within a query.  Query planning is difficult for protected search systems as they do not know statistics of the data.  Protected search generates a plan based on only query type.
\item The existence and complexity of rules (query policy and access control).  Protected search systems use advanced cryptography like multi-party computation to evaluate these rules, resulting in substantial overhead.
\end{enumerate}


\subsection{User Perceptions of Performance}
\label{sec:user-pilot}

We conducted a human-subjects usability evaluation to further the understanding of current protected search usability.
This evaluation considered the performance of multiple protected search technologies and the perception of performance by human subjects (our procedure was approved by our Institutional Review Board).  In this evaluation, subjects interacted with different protected search systems through an identical web interface.
Here, we focus on thoughts shared by participants during discussion.  (An informal overview of our procedure is in \apref{sec:pilot procedure}.)

%
%
Our participants discussed several themes that are salient for furthering the usability of protected search:

\begin{itemize}
\item Participants cared more about predictability of response times than minimizing the mean response time.  When response times were unpredictable, participants were unsure whether they should wait for a query to complete or do something else.
\item Participants felt the protected technologies were slower than an unprotected system.  Participants felt this performance was acceptable if it gave them access to additional data, but did not want to migrate current databases to a protected system.  Note that this feedback is from end users, not administrators.
\item Participants expected performance to be correlated with the number of records returned and the length of the query.  Participants were surprised that different types of queries might have different performance characteristics.
\end{itemize}

\subsection{Current Protected Search Databases}

Some protected search systems have made the transition to full database solutions.  These systems  report performance analysis, perform rule enforcement, and support dynamic data.

\begin{table*}[ht]
\centering
\begin{tabular}{|l|ccccccccc| p{.75cm} cc | cccc|c|c|}

&\rot{Equality}&\rot{Boolean}&\rot{Keyword}&\rot{Range}&\rot{Substring}&\rot{Wildcard}&\rot{Sum}&\rot{Join}&\rot{Update}&\rot{Approach}
&\rot{\# of parties}&\rot{Code available}
&\rot{Multi-client}&\rot{User auth.}&\rot{Access control}&\rot{Query policy}
&\rot{Leakage} & \rot{Performance}\\
System & \multicolumn{9}{c|}{Supported Operations} & \multicolumn{3}{c|}{Properties} & \multicolumn{4}{c|}{Features} & &\\
\hline

CryptDB~\cite{CACM:PRZB12} & \full & \full & \none & \full & \none & \none & \full & \full & \full
& \shortppe
& 2 & \full
& \full & \full & \full & \none
& \threeQuarter & \full \\
\hline

Arx~\cite{EPRINT:PodBoePop16} & \full & \none & \none & \full & \none & \none & \full & \full & \full
& \shortsse
& 2 & \none
& \none & \none & \none & \none
& \half & \half\\
\hline

BLIND SEER~\cite{SP:PKVKMC14,SP:FVKKKM15} & \full & \full & \full & \full & \none & \none & \none & \none & \full
& \shortsse
&3 & \none
& \full & \none & \none & \full
&\half & \half\\
\hline

OSPIR-OXT~\cite{C:CJJKRS13,CCS:JJKRS13,NDSS:CJJJKR14,ESORICS:FJKNRS15,ClusionPaper,clusion-github}  & \full & \full & \full & \full & \full & \full & \none & \none & \full
& \shortsse
& 3 & \none
& \full & \none & \none & \full
& \half & \threeQuarter\\
\hline

SisoSPIR~\cite{RSA:IKLO16} & \full & \none & \full & \full & \full & \none & \none & \none & \none
& \shortoram
&3 & \none
& \none & \none&\none&\full
& \none & \half\\
\hline

\end{tabular}
\vspace{.05in}
\caption{This table summarizes protected search databases that have been developed and evaluated at scale.  The \emph{Supported Operations} columns describe the queries naturally supported by each scheme.  \emph{Properties} and \emph{Features} columns describe the system and available functionality.  Finally \emph{Leakage} and \emph{Performance} describe the whole, complex system, and are therefore relative (vs. the more precisely defined values for individual operations used earlier).}
\label{tab:dbms}
\end{table*}

These systems are summarized in \tabref{tab:dbms}. CryptDB replicates most DBMS functionality with a performance overhead of under 30\%~\cite{CACM:PRZB12}.  This approach has been extended to NoSQL key-value stores~\cite{kepner2014computing,gadepally2015computing}. Arx is built on a NoSQL key-value store called mongoDB~\cite{mongodb}.  Arx reports a performance overhead of approximately $10\%$ when used to replace the database of a web application (ShareLatex). Blind Seer~\cite{SP:PKVKMC14} reports slowdown of between 20\% and 300\% for most queries, while OSPIR-OXT~\cite{C:CJJKRS13} report they occasionally outperform a baseline MySQL 5.5 system with a cold cache and are an order of magnitude slower than MySQL with a warm cache. The SisoSPIR system~\cite{RSA:IKLO16} reports performance slowdown of 500\% compared to a baseline MySQL system on keyword equality and range queries.

Given these performance numbers, we now ask which solution, if any, is appropriate for different database settings.

\subsubsection{Relational Algebra without Cartesian product}
CryptDB, Blind Seer, OSPIR-OXT, and SisoSPIR all provide functionality that supports most of relational algebra except for the Cartesian product operation.  These systems offer different performance/leakage tradeoffs.  CryptDB is the fastest and easiest to deploy.  However, once a column is used in a query, CryptDB reveals statistics about the entire dataset's value on this column.  The security impact of this leakage should be evaluated for each application (see \secref{sec:leakage-attack}).
Blind Seer and OSPIR-OXT also leak information to the server but primarily on data returned by the query.  Thus, they are appropriate in settings where a small fraction of the database is queried.  Finally, SisoSPIR is appropriate if a large fraction of the data is regularly queried.  However, SisoSPIR does not support Boolean queries, which is limiting in practice.

\subsubsection{Full Relational Algebra}
CryptDB is the only system for relational algebra that supports Cartesian product.  (As stated, while Kamara and Moataz~\cite{EPRINT:KamMoa16} support Cartesian product, but do not support dynamic data.)

\subsubsection{Associative Array - NoSQL Key-Value}
The Arx system built on mongoDB provides functionality necessary to support associative arrays.  
In addition, other commercial systems (e.g., Google's Encrypted BigQuery~\cite{encbigquery}) and academic works~\cite{kepner2014computing,gadepally2015computing} apply \shortppe techniques to build a NoSQL protected system.

Blind Seer, OSPIR-OXT, and SisoSPIR have sufficient query functionality to support associative arrays. However, their techniques concentrate on query performance. Associative array databases often have insert rates of over a million records per second.  The insert rates of Blind Seer, OSPIR-OXT, and SisoSPIR are multiple orders of magnitude smaller.  Suppose a record is being updated.  In an unprotected system this causes a small change to the primary index structure.  However in the protected setting, if only a few locations are modified the server may learn about the statistics of the updated record.  This creates a tension between efficiency and security.  Efficient updates are even more difficult if the provider does not have the full unprotected dataset.

\textbf{Open Problem:} Construct \shortsse and \shortoram techniques that can handle millions of inserts per second.

To support very large insert rates, NoSQL key-value stores commonly distribute the data across many machines.  This introduces the challenge of synchronizing queries, updates, and data across these machines.  This synchronization is difficult as none of the servers are supposed to know what queries, updates, or data they are processing!

\textbf{Open Problem:}
Construct protected search systems that leverage distributed server architectures.

\subsubsection{Linear Algebra and Others} No current protected search system supports the linear algebra basis used to implement complex graph databases.  In addition, as federated and polystore databases emerge it will be important to interface between different protected search systems that are designed for different query bases.

\emph{Inherent Limitations:} Protected search systems are still in development, so it is important to distinguish between transient limitations and inherent limitations of protected search.
Protected search inherently reduces data visibility in order to prevent abuse. To achieve high performance under these conditions, many design decisions such as the schema and the choice of which indices to build must be made before data is ingested and stored on the server. In particular, if an index has not been built for a particular field, then it simply cannot be searched without returning the entire database to the querier.  In general, it is not possible to dynamically permit a type of search without retrieving the entire dataset.

Additionally, if the database malfunctions, debugging efforts are complicated by the reduced visibility into server processes and logs. More generally, protected search systems are more complicated to manage and don't yet have an existing community of qualified, certified administrators.

Throughout this work we've identified a few transient limitations that can (and should!) be mitigated with future advances. 
Each potential user must make her own judgment as to whether the value of improved security outweighs the performance limitations.




\section{Conclusion and Outlook}
\label{sec:conclusion}
Several established and startup companies have commercialized protected search.  Most of these products today use the \shortppe technique, but we believe both \shortsse and \shortoram approaches will find their way into products with broad user bases.

\label{sec:pros cons}

Governments and companies are finding value in \emph{lacking} access to individuals' data~\cite{schneierdata}. Proactively protecting data mitigates the (ever-increasing) risk of server compromise, reduces the insider threat, can be marketed as a feature, and frees developers' time to work on other aspects of products and services.  The recent HITECH US Health Care Law~\cite{blumenthal2010launching} establishes a requirement to disclose breaches involving more than 500 patients but exempts companies if the data is encrypted: ``if your practice has a breach of encrypted data [...] it would not be considered a breach of
unsecured data, and you would not have to report it''~\cite{fedhealthcareprivacy}.

Protected database technology can also open up new markets, such as those cases where there is great value in recording and sharing information but the risk of data spills is too high
For example, companies recognize the value of sharing cyber threat and attack information~\cite{barnum2012standardizing}, but uncontrolled sharing of this information presents a risk to reputation and intellectual property.

This paper provides a snapshot of current protected search solutions.  There is currently no dominant  solution for all use cases.  Adopters need to understand system characteristics and tradeoffs for their use case.

Protected databases will see widespread adoption.  Protected search has developed rapidly since 2000, advancing from linear time equality queries on static data to complex searches on dynamic data, now within overhead between 30\%-500\% over standard SQL.

At the same time, the database landscape is rapidly changing, specializing, adding new functionality, and federating approaches.  Integrating protected search in a unified design requires close interaction between cryptographers, protected search designers, and database experts.
To spur that integration, we describe a three pronged approach to this collaboration: 1) developing base queries that are useful in many applications, 2) understanding how to combine queries to support multiple applications, and 3) rapidly applying techniques to emerging database technologies.

DBMSs are more than just efficient search systems; they are highly optimized and complex systems.  Protected search has shown that database and cryptography communities can work together.  The next step is to transform protected search systems into protected DBMSs.



\section*{Acknowledgments}

The authors thank David Cash, Carl Landwehr, Konrad Vesey, Charles Wright, and the anonymous reviewers for helpful feedback in improving this work.



\bibliographystyle{IEEEtran}
\bibliography{bibtex/noncrypto,bibtex/abbrev3,bibtex-arxiv/crypto,bibtex/vijay_trends,bibtex/missing-crypto}

\appendices

\section{Substring and Wildcard Query Combiners}
\label{sec:other combiners}
\setcounter{subsubsection}{\value{numcombiners}}
\subsubsection{Bounded-length substring using keyword equality \cite{RSA:IKLO16}}
Searches for substrings of a fixed length $\kappa$ can be supported simply by inserting all length-$\kappa$ substrings ($\kappa$-grams) into an equality-searchable index during initialization. 
Given a field with maximum length $\alpha$, this techniques requires adding $\alpha-\kappa$ keywords during insertion and making one keyword search during query execution.

\subsubsection{Short substring using range \cite{RSA:IKLO16}}
By inserting the $\kappa$-grams into a range index, queries for substrings of length up to $\kappa$ can also be supported. We explain by example: one can query for the two-character string ``hi'' by searching for the range $[hia, hiz]$ in an index of length-3 substrings.

\subsubsection{Anchored substring using conjunction \cite{C:CJJKRS13}}
We now consider the converse of the above situation: supporting searches of long substrings of length \emph{at least} $\kappa$, with storage overhead decreasing in $\kappa$.
We begin with an ``anchored'' search, where the substring occurs either at the beginning or end of the string.

By way of example, suppose we wish to support substring searches on the record $a=$``teststring''. In a conjunction-searchable index, we insert $\kappa$-grams of the string along with their location (1, ``tes''), (2, ``est''), \ldots, (8, ``ing'').  Now to search for all records containing ``test'' the client asks for all records matching both (1, ``tes'') and (2, ``est'').
Searching from the end of the string can be accomplished using negative indexing; using (-1, ``ing''), (-2, ``rin''), (-3, ``tri''), \ldots, (-8, ``tes'') in the above example.

\subsubsection{Substring using disjunction of conjunctions \cite{C:CJJKRS13}}
Removing the anchoring restriction from the above technique requires the use of disjunctions, since the starting location of the substring is unknown. To find the substring ``test'' the querier must search for a conjunction of ($i$, ``tes'') and ($i$+1, ``est'') for any starting position $i$. The number of terms in this formula depends on the maximum string length.

\subsubsection{ and 14) Wildcard using conjunctions \cite{C:CJJKRS13}}
The above technique also supports single-character wildcard queries. For instance, to search for ``tes\_str'', the client asks for a conjunction of (1, ``tes'') and (5, ``str'').  Note that the $\kappa$-gram length of letters is required on either side of the wildcard.  This technique can be extended for unanchored queries as above, and it supports multiple character wildcards by incrementing the expected positions of the $\kappa$-grams.


\section{Procedure for Pilot Study}
\label{sec:pilot procedure}

We installed and configured multiple protected search systems.
For each, we ingested ten million records of real application data, and conducted sessions with 10 users 
 over a ten-day period.  Our Institutional Review Board reviewed our protocols and questionnaires, determined that they 
represented a minimal risk, and approved the procedure.
Software for the procedure resided in an Amazon Web Services (AWS)~\cite{aws} network. 
Data was drawn from a genuine application source and was converted to a single, static table with over one hundred columns and ten million records.  

Participants had a mix of technical and non-technical backgrounds, with six men and four women.  All participants had prior experience interacting with web interfaces that use database backends to present results.
Participants were aware that they were using different systems but systems were identified only by a single letter.  
Participants were not given any information about the capabilities of the technologies. 

Each participant took part in three types of sessions, each of which lasted 30 minutes: 1) training on the web interface; 2) scripted interaction with each of the technologies; and 3) exploratory sessions with each of the technologies. 
Users interacted with the secure database technology through a web application which included a visual query builder which queried the underlying secure database. 
Participants interacted with the visual query builder to create queries. Then, the web server submitted the query to the protected search system.

\end{document}